\documentclass[prl,reprint,aps,superscriptaddress,nobibnotes,nofootinbib]{revtex4-2}

\usepackage{graphicx}
\usepackage{dcolumn}
\usepackage{bm}
\usepackage[tight]{units}
\usepackage[colorlinks,citecolor=blue,urlcolor=blue,linkcolor=blue]{hyperref}
\usepackage{hyperref}
\usepackage[normalem]{ulem}
\usepackage{verbatim}
\usepackage[dvipsnames]{xcolor}

\begin{document}


\title{Results on sub-GeV Dark Matter from a 10 eV Threshold CRESST-III Silicon Detector}%

\newcommand{\mpi}{\affiliation{Max-Planck-Institut f\"ur Physik, 80805 M\"unchen, Germany}}
\newcommand{\coimbra}{\affiliation{Also at: LIBPhys, Departamento de Fisica, Universidade de Coimbra, P3004 516 Coimbra, Portugal}}
\newcommand{\hephy}{\affiliation{Institut f\"ur Hochenergiephysik der \"Osterreichischen Akademie der Wissenschaften, 1050 Wien, Austria}}
\newcommand{\ati}{\affiliation{Atominstitut, Technische Universit\"at Wien, 1020 Wien, Austria}}
\newcommand{\tum}{\affiliation{Physik-Department, Technische Universit\"at M\"unchen, 85747 Garching, Germany}}
\newcommand{\tuebingen}{\affiliation{Eberhard-Karls-Universit\"at T\"ubingen, 72076 T\"ubingen, Germany}} 
\newcommand{\bratislava}{\affiliation{Comenius University, Faculty of Mathematics, Physics and Informatics, 84248 Bratislava, Slovakia}}

\newcommand{\oxford}{\affiliation{Department of Physics, University of Oxford, Oxford OX1 3RH, United Kingdom}}
\newcommand{\wmi}{\affiliation{Also at: Walther-Mei\ss ner-Institut f\"ur Tieftemperaturforschung, 85748 Garching, Germany}}
\newcommand{\lngs}{\affiliation{INFN, Laboratori Nazionali del Gran Sasso, 67010 Assergi, Italy}}
\newcommand{\gssi}{\affiliation{Also at: Gran Sasso Science Institute, 67100, L'Aquila, Italy}}
\newcommand{\cassino}{\affiliation{Also at: Dipartimento di Ingegneria Civile e Meccanica, Università degli Studi di Cassino e del Lazio Meridionale, 03043 Cassino, Italy}}

\mpi
\hephy
\ati
\lngs
\bratislava
\tum
\tuebingen
\oxford

\coimbra
\wmi
\gssi
\cassino

\author{G.~Angloher}
  \mpi

\author{S.~Banik}
  \hephy
  \ati

\author{G.~Benato}
  \lngs 

\author{A.~Bento}
  \mpi
  \coimbra 

\author{A.~Bertolini}
  \mpi

\author{R.~Breier}
  \bratislava

\author{C.~Bucci}
  \lngs 

\author{J.~Burkhart}
  \hephy

\author{L.~Canonica}
  \mpi 

\author{A.~D'Addabbo}
  \lngs

\author{S.~Di~Lorenzo}
  \lngs

\author{L.~Einfalt}
  \hephy
  \ati
  
\author{A.~Erb}
  \tum
  \wmi
  
\author{F.~v.~Feilitzsch}
  \tum 

\author{N.~Ferreiro~Iachellini}
  \mpi  
  
 \author{S.~Fichtinger}
  \hephy
 
\author{D.~Fuchs}
  \mpi  
 
\author{A.~Fuss}
  \hephy
  \ati

\author{A.~Garai}
  \mpi 
  
 \author{V.M.~Ghete}
  \hephy 

\author{S.~Gerster}
  \tuebingen 

\author{P.~Gorla}
  \lngs 

\author{P.V.~Guillaumon}
  \lngs

 \author{S.~Gupta}
  \hephy 

\author{D.~Hauff}
  \mpi
  \tuebingen

\author{M.~Ješkovsk\'y}
  \bratislava

\author{J.~Jochum}
  \tuebingen 

\author{M.~Kaznacheeva}
\email[Corresponding author: ]{margarita.kaznacheeva@tum.de}
  \tum

\author{A.~Kinast}
  \tum
  
\author{H.~Kluck}
  \hephy

\author{H.~Kraus}
  \oxford

\author{A.~Langenk\"amper}
  \tum
  \mpi

\author{M.~Mancuso}
  \mpi
 
 \author{L.~Marini}
  \lngs
  \gssi

\author{L.~Meyer}
  \tuebingen 
  
\author{V.~Mokina}
  \hephy
 
\author{A.~Nilima}
  \mpi 

\author{M.~Olmi}
  \lngs
  
\author{T.~Ortmann}
  \tum

\author{C.~Pagliarone}
  \lngs 
  \cassino

\author{L.~Pattavina}
  \tum
  \lngs

\author{F.~Petricca}
  \mpi 

\author{W.~Potzel}
  \tum 

\author{P.~Povinec}
  \bratislava

\author{F.~Pr\"obst}
  \mpi

\author{F.~Pucci}
  \mpi 
  
\author{F.~Reindl}
  \hephy
  \ati

\author{J.~Rothe}
  \tum
  
\author{K.~Sch\"affner}
  \mpi

\author{J.~Schieck}
  \hephy
  \ati 

\author{D.~Schmiedmayer}
   \hephy
   \ati

\author{S.~Sch\"onert}
  \tum 
  
\author{C.~Schwertner}
  \hephy
  \ati

\author{M.~Stahlberg}
  \mpi

\author{L.~Stodolsky}
  \mpi 

\author{C.~Strandhagen}
  \tuebingen

\author{R.~Strauss}
  \tum

\author{I.~Usherov}
  \tuebingen 

\author{F.~Wagner}
  \hephy

\author{M.~Willers}
  \tum 

\author{V.~Zema}
  \mpi

\collaboration{CRESST Collaboration}
\noaffiliation


\begin{abstract}
We present limits on the spin-independent interaction cross section of dark matter particles with silicon nuclei, derived from data taken with a cryogenic calorimeter with \unit[0.35]{g} target mass operated in the CRESST-III experiment. A baseline nuclear recoil energy resolution of ${\unit[(1.36\pm 0.05)]{eV_{nr}}}$, currently the lowest reported for macroscopic particle detectors, and a corresponding energy threshold of ${\unit[(10.0\pm 0.2)]{eV_{nr}}}$ have been achieved, improving the sensitivity to light dark matter particles with masses below \unit[160]{MeV/c$^2$} by a factor of up to 20 compared to previous results. We characterize the observed low energy excess, and we exclude noise triggers and radioactive contaminations on the crystal surfaces as dominant contributions. 
\end{abstract}

\maketitle


\section{\label{sec:intro}Introduction}
Dark matter (DM) is the prevailing hypothesis that explains a variety of observed phenomena in the universe~\cite{PDG}. There is currently a wealth of well-founded theories proposing new types of particles that make up this DM~\cite{BAER20151LDM,Gelmini_2017LDM,Roszkowski_2018DM, Silk}. A wide range of DM masses could be realized in nature, motivating a broad search strategy with different experiments probing various DM models. In direct detection experiments the scattering of DM particles with a target is tested~\cite{goodman1985}. Since electromagnetic interaction must be strongly suppressed or non-existent for DM particles, scattering would primarily happen on the target nuclei. For a direct detection experiment to be sensitive to nuclear recoils of DM particles with masses of ${\unit[1]{GeV}/\text{c}^2}$ and smaller, extremely low energy thresholds of $\mathcal{O}$(\unit[10]{eV}) are required. Since DM particles have not yet been detected, results are usually presented in terms of upper limits on the cross section of the DM-nucleon interaction for a range of DM particle masses.

The Cryogenic Rare Event Search with Superconducting Thermometers (CRESST) is a DM direct detection experiment that uses crystals as cryogenic calorimeters in the low background environment of the deep underground facility of the Laboratori Nazionali del Gran Sasso (LNGS) in Italy~\cite{detA}. The third phase of the CRESST experiment, CRESST-III, focuses on low-mass DM. By using tungsten Transition Edge Sensors (TES)~\cite{cabrera2008TES}, CRESST achieves energy thresholds of $\mathcal{O}$(\unit[10]{eV}) and is one of the leading experiments in the search for DM particles in the sub-GeV range. Another advantage of the CRESST technology is the flexibility in the choice of target materials: the first results on low-mass DM with CRESST-III were obtained with a scintillating CaWO$_4$ crystal~\cite{detA}, which provides excellent particle discrimination, while the use of LiAlO$_2$ as a target significantly improved the results on spin-dependent interactions of DM with protons and neutrons~\cite{Li_paper}.

After the part of the energy spectrum below several hundred eV got accessible, sharply rising rates of events towards the threshold have been measured in all CRESST-III detectors~\cite{detA,Li_paper,LEEpaper}. This observation is often referred to as low energy excess (LEE) due to the fact that the number of registered events highly exceeds the values expected from known background sources. Similar features are observed in other low-threshold cryogenic experiments, such as EDELWEISS~\cite{edelweissSurf2019,edelweissUG2020,edelweissNbSi2022}, NUCLEUS~\cite{nucleus_prototype,rothe2020nucleus}, SuperCDMS-CPD~\cite{cpd}, SuperCDMS-HVeV~\cite{SuperCDMSHveVDM_2020,SuperCDMS_HVeV_excess2022}. These excesses are currently the main limiting factor for further sensitivity improvement of cryogenic DM searches and are therefore intensively studied in the community~\cite{ExcessData}. A recent overview of experimental observations of the excesses in the experiments listed above as well as experiments exploiting charge readout via CCD-sensors, DAMIC~\cite{damic2020} and SENSEI~\cite{sensei2020}, can be found in Ref.~\cite{ExcessWorkshop}. 

In this article, we present the DM search results of a recent data-taking campaign, between November 2020 and August 2021, with a cryogenic Si detector. First, we provide details on the experimental setup and detector design in Sec.~\ref{sec:setup}. Then, in Sec.~\ref{sec:analysis}, we describe the data analysis procedure and report the detector performance. In Sec.~\ref{sec:spectrum}, we show and discuss the measured energy spectrum and constrain the possible origins of the observed LEE. Finally, we explore new parameter space for low-mass DM particles with our data in Sec.~\ref{sec:DMresults}.

\section{\label{sec:setup}Experimental setup}
The CRESST-III experimental setup is located at LNGS and is protected by \unit[3600]{m.w.e.} rock overburden, which provides excellent shielding against cosmic rays. In particular, the muon flux is suppressed by six orders of magnitude compared to sea level~\cite{Bellini_2012_lngs_borexino}. In addition, the detectors are protected by concentric layers of shielding materials. The outer layer of polyethylene thermalizes environmental neutrons. The following layer of plastic scintillators acts as an active muon veto and tags remaining cosmic muons. An airtight nitrogen-purged box is installed inside the muon veto to avoid the accumulation of radon gas and its decay products. Inside the box, there are layers of low-radioactivity lead and copper that shield against $\gamma$ backgrounds. An additional thin inner layer of polyethylene shields the cryogenic detectors, which are operated in a $^3$He/$^4$He dilution refrigerator, from neutrons emitted by the lead and copper shielding. More details on the experimental setup can be found in~Ref.~\cite{cresst2012results}.

CRESST-III detectors consist of CaWO$_4$, Al$_2$O$_3$, Si or LiAlO$_2$ single crystals operated as cryogenic calorimeters at temperatures around \unit[15]{mK} (see Ref.~\cite{LEEpaper} for a recent overview). A particle recoil in the target crystal produces phonons. In a good approximation for non-scintillating calorimeters, the total energy transferred to the phonon system corresponds to the energy deposit, independently of the type of interacting particle, i.e, nuclear or electron recoil, over a wide energy range.
Thus, for the Si detector presented in this work, we assume the energy scale independent of the type of interaction. The phonons are collected by a TES sensor~\cite{cabrera2008TES} with Al phonon collectors directly evaporated on the crystal and cause a temperature rise in the tungsten film. The TESs are operated at a temperature between the tungsten super- and normal conducting phases, which leads to strong responses to temperature fluctuations down to $\mathcal{O}(\mu \text{K})$. A readout circuit based on a SQUID amplifier finally produces a voltage signal proportional to the heat fluctuations in the TES~\cite{cresst2012results}. 

\begin{figure}[t!]
\includegraphics[width=\linewidth]{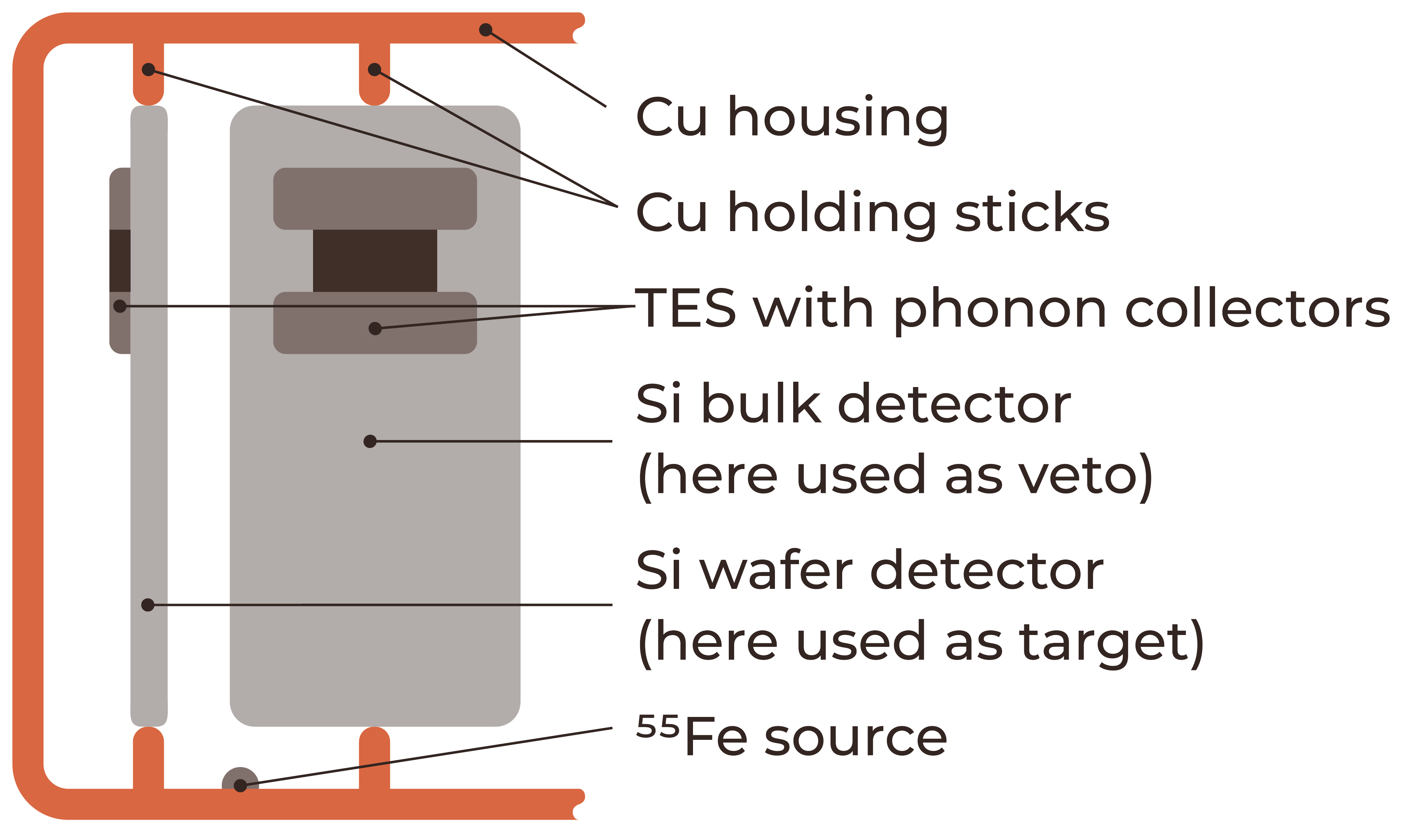}
\caption{\label{fig:Design} Schematic drawing of the Si CRESST-III detector module. In this work, we use the Si wafer detector as the main target crystal, while the bulk Si crystal is used as a veto detector. The crystals are equipped with TES, held by three Cu sticks, and mounted inside a Cu housing. The  $^{55}$Fe calibration source encapsulated with concentric layers of glue and gold is attached to the module's wall. The drawing is not to scale.}
\end{figure}

In this work, we focus on a light 0.35~g thin wafer-like Si detector with the size of \unit[($20\times20\times0.4$)]{mm$^3$}. The crystal is instrumented with an individual TES and held by three Cu sticks. One side of this wafer is facing a second instrumented Si crystal with the size of \unit[($20\times20\times10$)]{mm$^3$} enabling the study of coincident events. Both are encapsulated in a Cu housing. Fig.~\ref{fig:Design}  shows a schematic drawing of this module comprising the aforementioned components. In contrast to the standard CRESST-III design~\cite{detA}, no scintillation light is produced within this module. We have chosen to use the wafer detector as the main target due to its superior performance in comparison to the bulk detector.

 The detector is equipped with an ohmic heater (thin Au film) to heat the TES into the operating point in the transition and inject artificial pulses. The latter are used to monitor the detector's stability and response. A feedback loop corrects changes in the measured height of the artificial pulses via the heater in order to ensure that the detector stays in its operating point~\cite{CRESST2009Commissioning}.

An absolute energy calibration is obtained by a low activity (${\sim \unit[1]{mBq}}$) $^{55}$Fe source attached to one of the walls of the module as shown in Fig.~\ref{fig:Design}. The source is covered by a layer of glue to suppress the Auger electrons reaching the crystal. A layer of gold is added on top of the structure to shield the detectors from possible scintillation light created in the glue.

\section{\label{sec:analysis}Data processing and analysis}

In this work, we use data from 186.9~days of measurement. We first adjusted all the steps described below on the training data set of 29.6~days of measurement, while the remaining data equivalent to 157.3~days was blinded. After the analysis procedure was fixed, we applied it without changes to the blinded data set. 
    
In CRESST-III, data are recorded continuously with a sampling frequency of \unit[25]{kHz}. We perform offline triggering after filtering the stream with an optimum filter~\cite{OF}. To calculate the optimum filter two components are required: the typical noise power spectrum (NPS) of the detector and the expected pulse shape of particle events. The former we calculate by averaging the NPS of a large number of randomly selected noise traces. To obtain the pulse template we first sum up several hundred valid registered events from the linear range of the detector response which extends up to \unit[300]{eV}. Then we fit the result with a well-established pulse shape model described in Ref.~\cite{probst1995model} to fully remove noise from the template. We store triggered events in windows with a length of 16384 data samples corresponding to \unit[655.36]{ms} for further analysis. 
    
For this detector we set the trigger threshold to a value of \unit[12.8]{mV} allowing only one noise trigger per kg-day of exposure following the  procedure described in Ref.~\cite{trigger}. The procedure uses a noise trigger model to provide a value for the threshold as a function of the expected number of noise triggers. The expected noise trigger rate is discussed in Sec.~\ref{sec:spectrum}. 
    
We use the optimum filter method for amplitude reconstruction in the linear regime of the detector response. If the energy deposition in the crystal is large enough for the TES to get close to the flat part of the superconducting transition curve, the detector response is not linear anymore. Therefore, the pulse shape is distorted and hence amplitude reconstruction with the optimum filter cannot be used anymore. For pulses outside of the linear regime, we use a truncated template fit described, e.g., in Ref.~\cite{Stahlberg_thesis}.

\begin{figure}[t]
\centering
   \includegraphics[width=\linewidth]{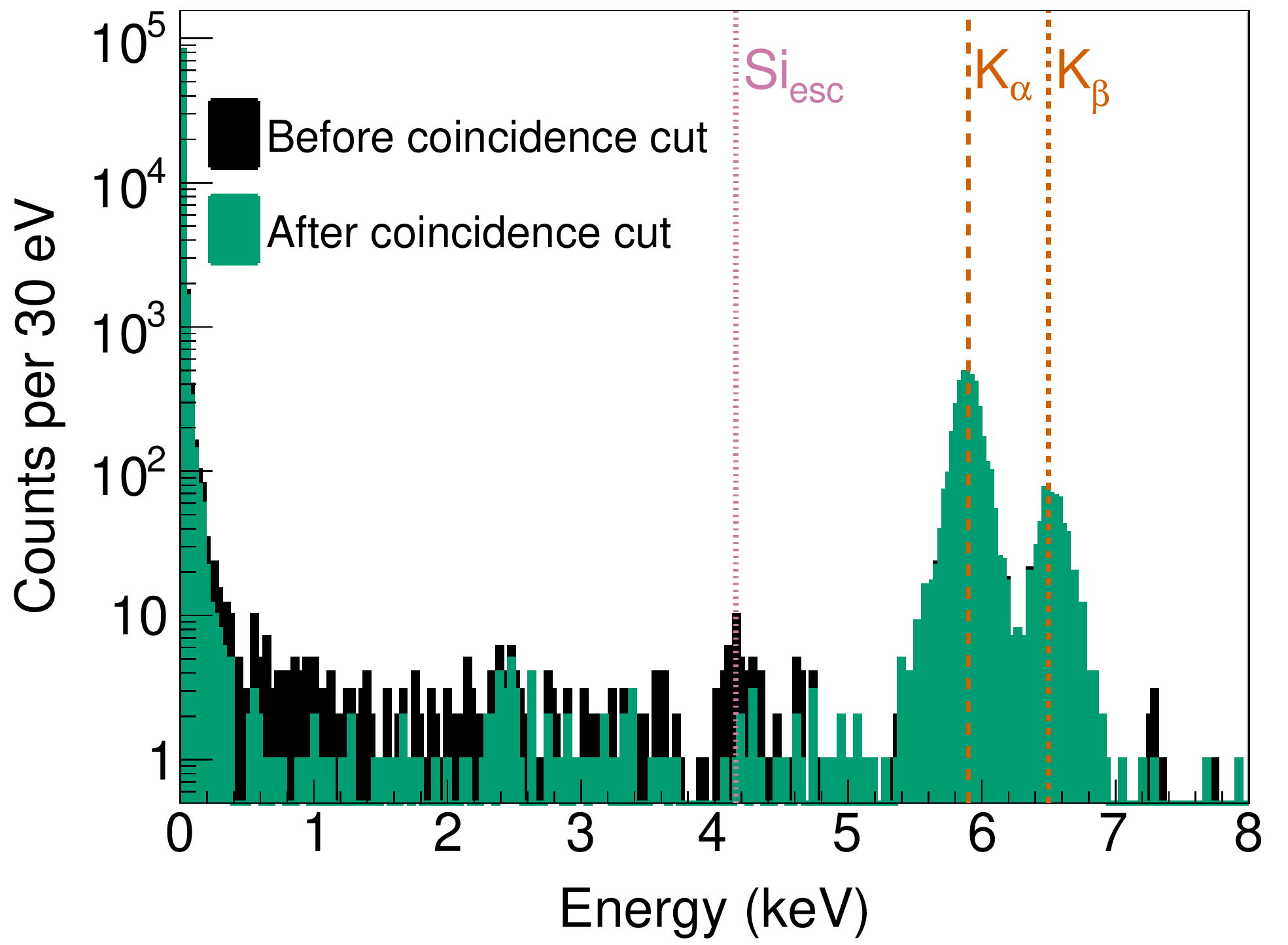}
  \caption{Measured energy spectrum of the Si wafer detector before the coincidence cut with the bulk detector facing the wafer (black) and after the coincidence cut (green). The dashed orange lines indicate the $^{55}$Mn K$_{\alpha}$ (\unit[5.9]{keV}) and K$_{\beta}$ (\unit[6.5]{keV}) lines from the source used for the energy calibration. The dotted pink line shows the Si escape line from the K$_{\alpha}$ X-ray at \unit[4.16]{keV}. No correction with the trigger and cut efficiencies is applied (see text for details).}
  \label{fig:antiCoinc}
\end{figure}

The electron capture in the $^{55}$Fe source provides $^{55}$Mn K$_\alpha$ and K$_\beta$ X-ray lines at 5.9 and \unit[6.5] {keV} respectively~\cite{xrays} used for energy calibration. They are marked with orange lines in Fig.~\ref{fig:antiCoinc}. For a precise calibration over a wide energy range, we use a set of artificial pulses of fixed discrete energies extending from threshold up to \unit[17]{keV} sent periodically every 20 seconds via the Au heater. Since the amplitude of a pulse is a measure of the energy injected to the crystal, we use these artificial pulses to map the detector response over the entire energy range. Additionally, this allows correcting possible detector response drifts over time~\cite{Stahlberg_thesis}.

We select valid events not caused or affected by artifacts or detector instabilities by applying a set of quality cuts. These cuts are based on features of the pulse shape and baseline, and follow the objective to reject events that deviate from the particle pulse shape. Additionally, we exclude periods of unstable detector operation by excluding time intervals where the height of the injected control pulses differs from the desired value by more than two standard deviations.
 
To obtain a data set for the final DM search we apply coincidence cuts, since a DM particle is not expected to scatter in multiple detectors simultaneously. We remove all pulses with a coinciding signal in any other operated detector module within a time window of ${\pm \unit[10]{ms}}$, all pulses for which at least one of the muon veto panels triggered within ${\pm \unit[5]{ms}}$. Additionally, we remove all coincidences within a record window with the Si bulk detector located in the same module. The final energy spectrum of the Si detector, before and after application of the coincidence cut with the bulk detector, is shown in Fig.~\ref{fig:antiCoinc} and discussed in Sec.~\ref{sec:spectrum}.

To evaluate the probability of a valid event passing all selection criteria described above, we applied the same analysis scheme to $3.8\times10^6$ simulated events. These events have the pulse shape of the particle template and energies uniformly distributed between 0 and \unit[300]{eV} (upper limit of the linear detector response). The simulated events are superimposed on the real data stream at random times. 

\begin{figure}[t]
\centering
   \includegraphics[width=\linewidth]{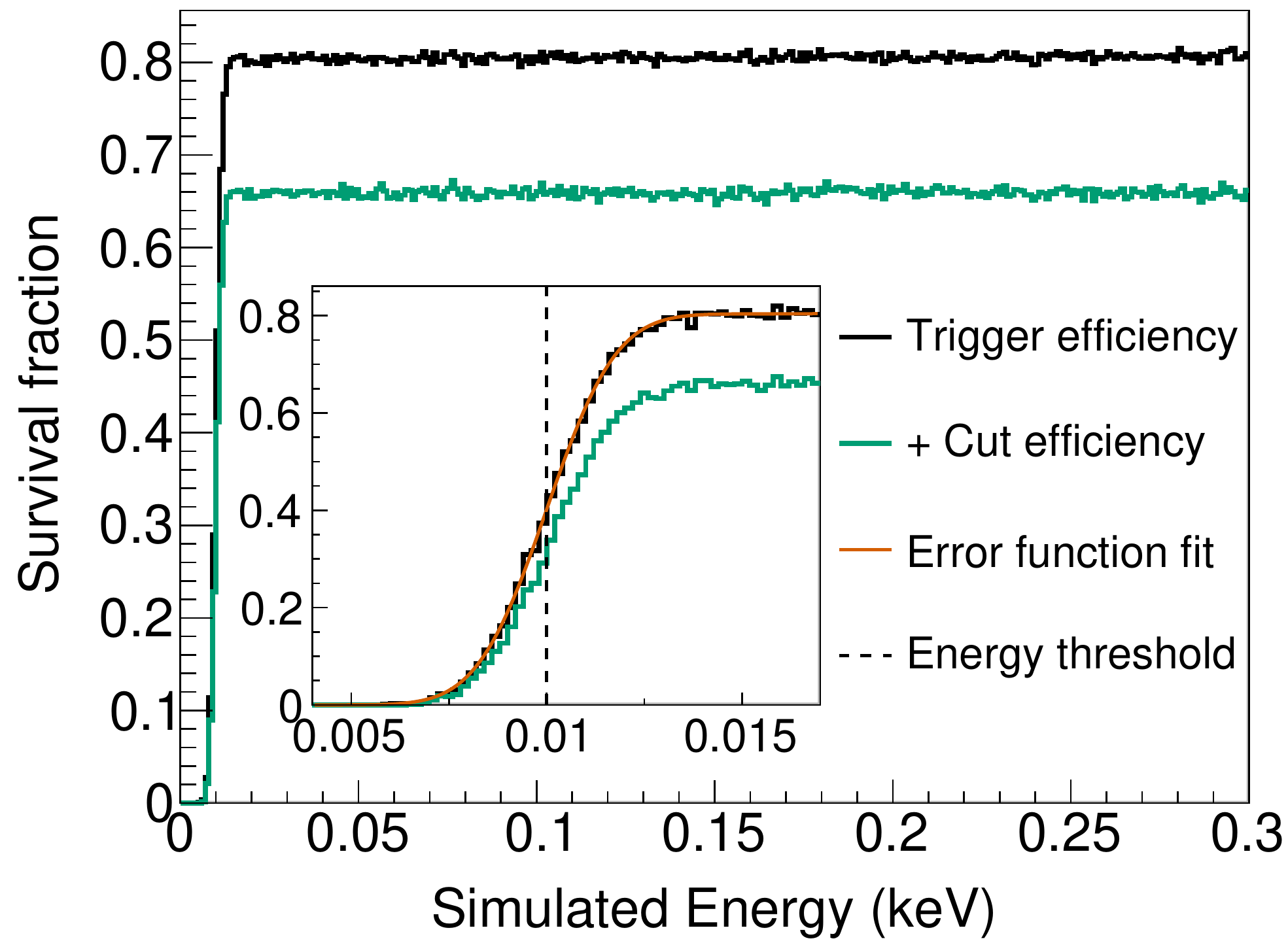}
  \caption{The fraction of simulated events that caused a trigger (black) and passed all selection criteria (green). The inset shows a zoom into the low energy part where the trigger probability is highly energy dependent. The solid orange line shows the error function fit that gives an energy threshold of ${\unit[(10.0\pm 0.2)]{eV}}$ (dashed black line) and baseline resolution of ${\unit[(1.36\pm 0.05)]{eV}}$ for the Si wafer detector.}
  \label{fig:eff}
\end{figure}

The fraction of simulated events causing a trigger is shown in black in Fig.~\ref{fig:eff}. Its flat part corresponds to a survival fraction of $(80.53\pm0.02)\%$ where the remaining events are lost due to the trigger dead time caused by coincidences with heater pulses or real particle events. 

The trigger efficiency decreases close to the threshold. After removing pile-ups we define the energy threshold at the level where 50\% of the simulated pulses which do not fall into the the dead time are triggered (dashed line in the inset of Fig.~\ref{fig:eff}). This corresponds to $\text{40.3\%}$ of the trigger efficiency. The fit of an error function to the trigger efficiency results in an energy threshold of ${E_{\text{thr}}={\unit[(10.0\pm 0.2)]{eV}}}$, a detector baseline resolution of ${\sigma_{\text{BL}}=\unit[(1.36\pm 0.05)]{eV}}$, and is shown as the orange line. 

We cross-check the fitted values with alternative methods to estimate the energy threshold and resolution. We convert the nominal trigger threshold voltage value to energy units, which gives ${\unit[(10.24\pm 0.15)]{eV}}$. We obtain the baseline resolution by superimposing a noiseless pulse template to a set of noise traces randomly collected over the measuring time. In this case, the amplitude reconstruction is only distorted by the baseline noise fluctuations and thus gives a measure for the baseline resolution, i.e. the resolution at zero energy, of ${\unit[(1.35\pm 0.02)]{eV}}$. The latter values agree with the ones from the error function fit within the uncertainties and thus additionally support their robustness.

Next, we apply the selection criteria to the triggered simulated events to evaluate the fraction of valid pulses that survive our analysis chain. To avoid an overestimation of the survival probability due to misidentified events, we remove all outliers where the reconstructed and simulated energies differ by more than three baseline resolution values. This selection implies that events coinciding with a noise fluctuation higher than ${\unit[3]{\sigma_{BL}}}$ are removed. This treatment of sub-threshold simulated events defines the lower mass limit for DM search and, thus, makes the DM results more conservative. The last explained selection is included in the cumulative probability for a simulated event with a given energy to be triggered and pass all the cuts shown in Fig.~\ref{fig:eff} as the green line. Above \unit[14]{eV} the survival probability remains constant at the value of ${(65.91\pm0.03)\%}$ which indicates that the cuts applied do not introduce any significant energy dependence.

\section{\label{sec:spectrum}Observed energy spectrum}

We obtained values for the energy threshold ${E_{\text{thr}}={\unit[(10.0\pm 0.2)]{eV}}}$ and the detector baseline resolution ${\sigma_{\text{BL}}=\unit[(1.36\pm 0.05)]{eV}}$ (see Sec.~\ref{sec:analysis}). This is a large improvement in comparison to the previous best performance in CRESST of \unit[30.1]{eV} for the threshold and \unit[4.6]{eV} for the detector resolution~\cite{detA}. Moreover, this is the best nuclear recoil energy resolution achieved with a macroscopic particle detector reported to date. On the one hand, this excellent performance allows probing DM particles with smaller masses. On the other hand, it opens the possibility to extend the range for studying the LEE. In this section, we first discuss the measured energy spectrum and the impact of the coincidence cut with the bulk detector. After that, we focus on the low energy part of the spectrum.

The resulting recoil energy spectrum after applying all the steps described in Sec.~\ref{sec:analysis} is shown in black in Fig.~\ref{fig:antiCoinc}, where several features are prominent. The spectrum is shown without applying the trigger and cut efficiencies which are roughly constant at the energy range above \unit[14]{eV} and thus do not change the spectral shape. With the dashed orange lines we indicate the $^{55}$Mn K$_{\alpha}$ \unit[5.9]{keV} and K$_{\beta}$ \unit[6.5]{keV} lines from the $^{55}$Fe source used for the energy calibration. The Si escape line from the K$_{\alpha}$ X-ray is shown with the dotted pink line. This process occurs when an X-ray from the calibration source excites a Si atom in the target crystal and the Si X-ray escapes the detector. Thus, the expected energy deposition is \unit[4.16]{keV}, which corresponds to the difference between the K$_{\alpha}$ energy from the source and the Si X-ray (\unit[1.74]{keV}). We measure this line at ${\unit[(4.18\pm 0.02)]{keV}}$, which is in agreement with the expected value and provides a cross-check for the calibration procedure at lower energies. This line is significantly reduced after removing the events that have a coinciding energy deposition in the bulk detector facing the wafer. This is expected due to a high probability of catching the escaping Si X-ray with the bulk detector. The Si X-ray line cannot be found in the energy spectrum of the wafer detector because the source is not positioned in the center of the gap between the wafer and bulk crystals (see Fig.~\ref{fig:Design}) and thus the solid angles covered by the two crystal sides facing each other are significantly different (confirmed by simulations). Furthermore, we observe a peak structure in the energy spectrum after the coincidence cut at ${\unit[(2.4\pm 0.1)]{keV}}$. This energy value suggests a scenario where two Si X-rays escape the crystal after a $^{55}$Mn K$_{\alpha}$ X-ray is absorbed by the wafer with the expected remaining energy deposition to the crystal of \unit[2.42]{keV}. Simulation studies are ongoing to clarify whether such process could be possible, e.g., at the edges of the wafer or on rough surfaces. Although the origin of this structure remains under investigation, its presence does not affect the results presented in this work since it appears far beyond the region of interest of our studies.

\begin{figure}[t]
\centering
   \includegraphics[width=\linewidth]{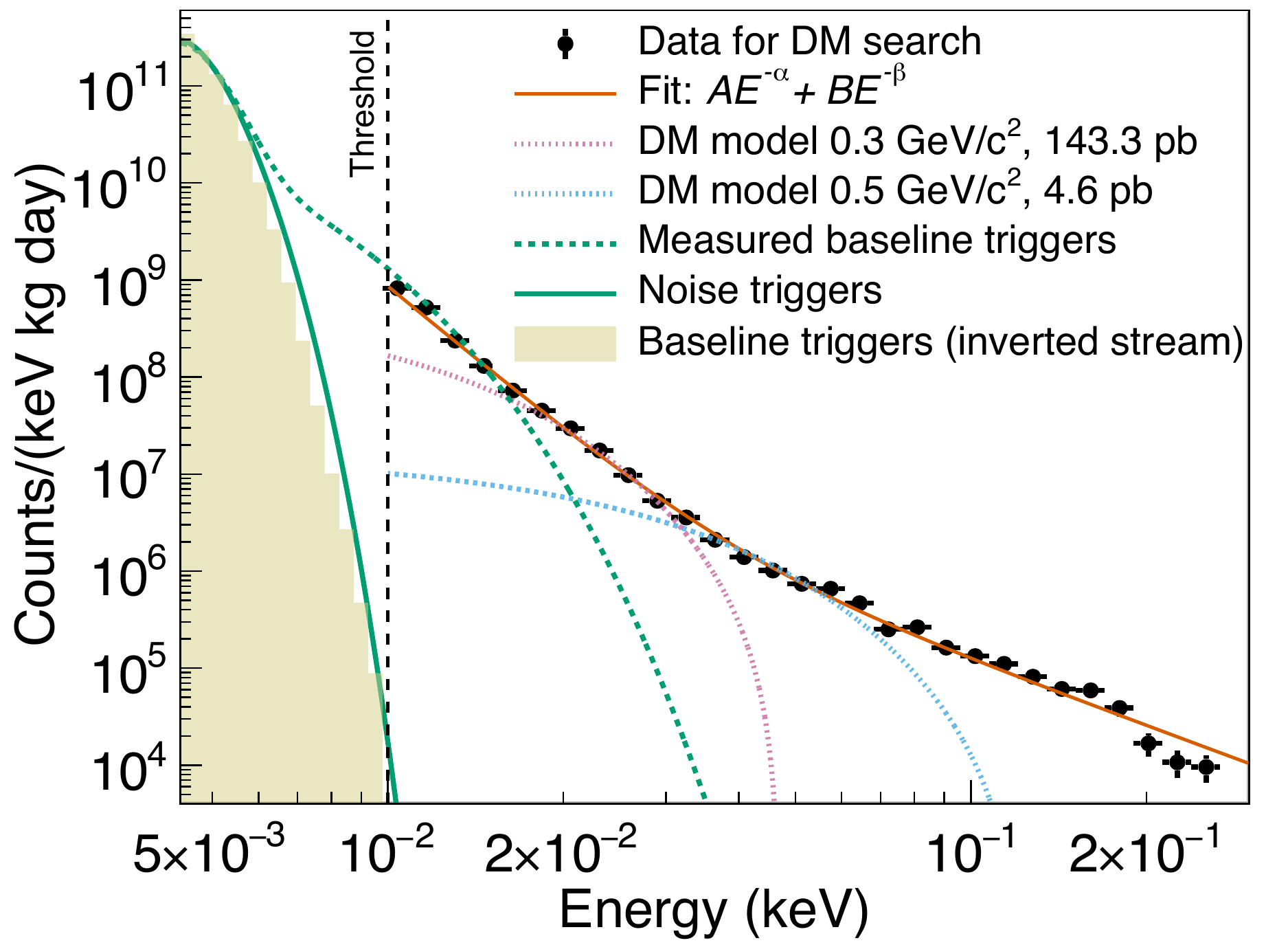}
  \caption{The energy spectrum (black data points) below \unit[300]{eV} after applying the selection cuts for the DM analysis taking into account the trigger and cut efficiencies. It exhibits a strong rise in the event rate towards lower energies, the LEE. We fit the LEE with the sum of two power laws (Eq.~\ref{eq:fit}, orange line). The dashed green line shows the distribution of the baseline triggers  derived from analyzing the randomly collected frames (see text for details). The solid green line shows the distribution of the noise triggers. A gap between this distribution and the data rules out noise triggers as the source of the LEE. The yellow histogram shows the distribution obtained from applying the optimum filter to the inverted data stream perfectly agreeing with the noise trigger distribution. Additionally, the differential energy spectra for two DM masses, \unit[0.3]{GeV/c$^2$} (pink) and \unit[0.5]{GeV/c$^2$} (blue), expected for interaction cross sections excluded with {90\%~C.L.} in this work (red limit line in Fig.~\ref{fig:limits}) are shown.}
  \label{fig:LEEfit}
\end{figure}

The coincidence cut also removes the background events originating from contaminations of crystal surfaces and events introduced by the presence of the calibration source in the module. In the energy range of \unit[(0.5--2)]{keV} a significant reduction of the event rate, by ${(76 \pm 5)}$\%, is observed. To understand the observed coincident events that originate from the calibration source, we performed a Monte Carlo simulation with the source placed inside the module as shown in Fig.~\ref{fig:Design}. For this we used the ImpCRESST physics simulation code~\cite{abdelhameed2019geant4, abdelhameed2019err} based on Geant4~\cite{agostinelli2003geant4,Allison2006,Allison2016} version 10.06 patch 3. The results suggest that a large share of coinciding events is expected to stem from multiple scattering processes of calibration source X-rays. This is supported by the results from an Al$_2$O$_3$ detector operated with a removable $^{55}$Fe source described in~\cite{phdthesis_rothe}. We assume that the rest of the events removed by the coincidence cut are radioactive backgrounds accumulated on the crystal surfaces. 

Another prominent feature in the energy spectrum (Fig.~\ref{fig:antiCoinc}) is the sharp increase of the event rate below \unit[300]{eV}. This observation was first made in CRESST-III~\cite{detA} when the energy threshold was lowered to \unit[30]{eV}. Since then we have observed this feature in all low-threshold CRESST-III detector modules with different target materials and geometries. A detailed investigation of the LEE based on the measurements from the recent CRESST-III data-taking campaign (including the data discussed in this work) is presented in Ref.~\cite{LEEpaper}. Even though the origin of those events so far remains unknown, DM interactions as well as intrinsic and external radioactive backgrounds as a major contribution to the LEE are excluded~\cite{LEEpaper}. The coincidence cut with the bulk detector removes only a negligible ${(0.9\pm 0.5)}$\% share of events below \unit[300]{eV} and therefore we can  exclude the crystal-surface background as dominant origin of the LEE. Additionally, in contrast to the previous CRESST-III module design~\cite{detA}, no scintillating materials are present in the detector module discussed in this work. Thus we also rule out the scintillation light as a significant contribution to the LEE. The observed increase of the LEE rate after warming up the cryostat to \unit[60]{K} reported in Ref.~\cite{LEEpaper} suggests a large impact of solid state physics effects activated in the target crystals or their interfaces with the holding structures or TES.

\begin{table}[!t]
\centering
\begin{tabular}{@{}cc@{}} \toprule
Fit parameter & Value \\
\hline
\bf{${A}$} &  $(7.6 \pm 2.0) \cdot 10^{-2} (\text{keV}^{(1 - \alpha)} \cdot \text{kg} \cdot \text{day})^{-1} $\\
$\alpha$ & $5.02 \pm 0.06 $  \\
$B$ &  $(7.2 \pm 1.5) \cdot 10^{2} (\text{keV}^{(1 - \beta)} \cdot \text{kg} \cdot \text{day})^{-1} $ \\
$\beta$ & $2.22 \pm 0.09$ \\
\hline
\hline
\end{tabular}
\caption{The values of the free fit parameters obtained from a $\chi^2$ fit of Eq.~\ref{eq:fit} to the spectrum shown in~Fig.~\ref{fig:LEEfit}. While the fitting function describes the shape of the LEE well, we do not associate a physical meaning with it. It provides a means of comparison with other experiments exhibiting excesses at low energies.}
\label{tab:LEE_fit}
\end{table}

We show the energy spectrum after cuts below \unit[300]{eV} converted to rate by taking into account the trigger and cut efficiencies as black data points in Fig.~\ref{fig:LEEfit}. The error bars represent the statistical uncertainties. Similarly to the procedure described in Ref.~\cite{trigger}, we apply the optimum filter to the set of baseline traces randomly collected from the data stream and derive the expected distribution of the energies of these baseline triggers shown as the dashed green line in Fig.~\ref{fig:LEEfit}. The right shoulder of this distribution matches the energy spectrum measured close to threshold very well. We expect this effect, as many of the randomly collected baselines are in coincidence with small pulses from the LEE. To account for the contribution from small, polluting pulses we extended the baseline trigger model described in Ref.~\cite{trigger} with an exponentially distributed pollution component associated with the LEE. 

The remaining Gaussian component of the baseline trigger model forms the detector noise distribution and is shown with the solid green line. We use it to apply the threshold determination method described in Ref.~\cite{trigger} to allow only one noise trigger per kg-day of exposure. It leads to a rather conservative threshold at the level of ${\unit[7.35]{\sigma_{BL}}}$. As can be seen in Fig.~\ref{fig:LEEfit}, the contribution of the noise triggers to the LEE rate above threshold is negligible. 

As an additional confirmation of our noise trigger spectrum, we apply the optimum filter to the voltage-inverted data stream. Since the trigger algorithm only searches for positive voltage signals and real temperature variations are always positive, any detector signal is not found anymore after the inversion. Only noise signals fluctuate symmetrically in both positive and negative direction and, hence, only they are picked up by the algorithm and are the sole contribution to the energy spectrum of the inverted stream. These upward fluctuations found in the inverted data stream are shown as the yellow histogram in~Fig.~\ref{fig:LEEfit}. This distribution was scaled with the exposure of the inverted data and thus its agreement with the solid green line confirms the validity of the noise trigger model.

To enable a quantitative comparison of the shape of the observed excesses among different detectors and experiments~\cite{ExcessWorkshop}, we introduce a fit function for the low-energy data. This function does not have an immediate physical motivation and only describes the spectrum in the energy region relevant to the LEE. It is defined as a sum of two power-law functions\footnotemark \addtocounter{footnote}{-1}
\footnotetext{The fit function we chose differs from the one described in Ref.~\cite{ExcessinSemiconductors2022}. We do not include an exponential component since we have excluded noise triggers from the data as shown above.}

\begin{equation}\label{eq:fit}
    f(E) = AE^{-\alpha} + BE^{-\beta}, 
\end{equation}
where $E$ is the energy and $A$, $B$, $\alpha$, $\beta$ are free fit parameters. Tab.~\ref{tab:LEE_fit} details the values obtained using a $\chi^2$ fit to the binned spectrum below \unit[300]{eV}. The fit is shown as the orange line in Fig. \ref{fig:LEEfit}.

\section{\label{sec:DMresults}Dark matter results}

We conservatively consider all events that survived the selection criteria described in Sec.~\ref{sec:analysis} with $E>E_{\text{thr}}$ to be potential DM signals originating from elastic DM-nucleus scattering (black data points in Fig.~\ref{fig:LEEfit}). The gross exposure of the blind data set used for the DM search is 55.06~g-day. Since the precise estimation of the trigger and cut survival probability can only be performed for the linear range of the detector response below \unit[300]{eV}, we exclude events outside this range from the DM search. The spectrum of the resulting region of interest is shown in black in Fig.~\ref{fig:LEEfit}.

For each DM mass the expected differential energy spectrum can be calculated under the assumptions of the standard DM halo model~\cite{DMhaloModel}: a local DM mass density ${\rho_{\text{DM}}=\unit[0.3]{(GeV/c^2)/cm^3}}$~\cite{salucci2010dark}, a Maxwellian velocity distribution with an asymptotic velocity ${v_\odot=\unit[220]{km/s}}$~\cite{Kerr} and a galactic escape velocity ${v_{\text{esc}}=\unit[544]{km/s}}$~\cite{Smith2007}. In case DM particles with a given mass interact with the detector material this expected recoil energy spectrum would be observed with distortions from the finite energy resolution of the detector, the trigger algorithm and analysis chain. We estimate this cumulative effect from the simulations described in Sec.~\ref{sec:analysis}. Thus, we can calculate how an energy spectrum for a given DM mass would look when measured by our detector.

\begin{figure}[!t]
\centering
   \includegraphics[width=\linewidth]{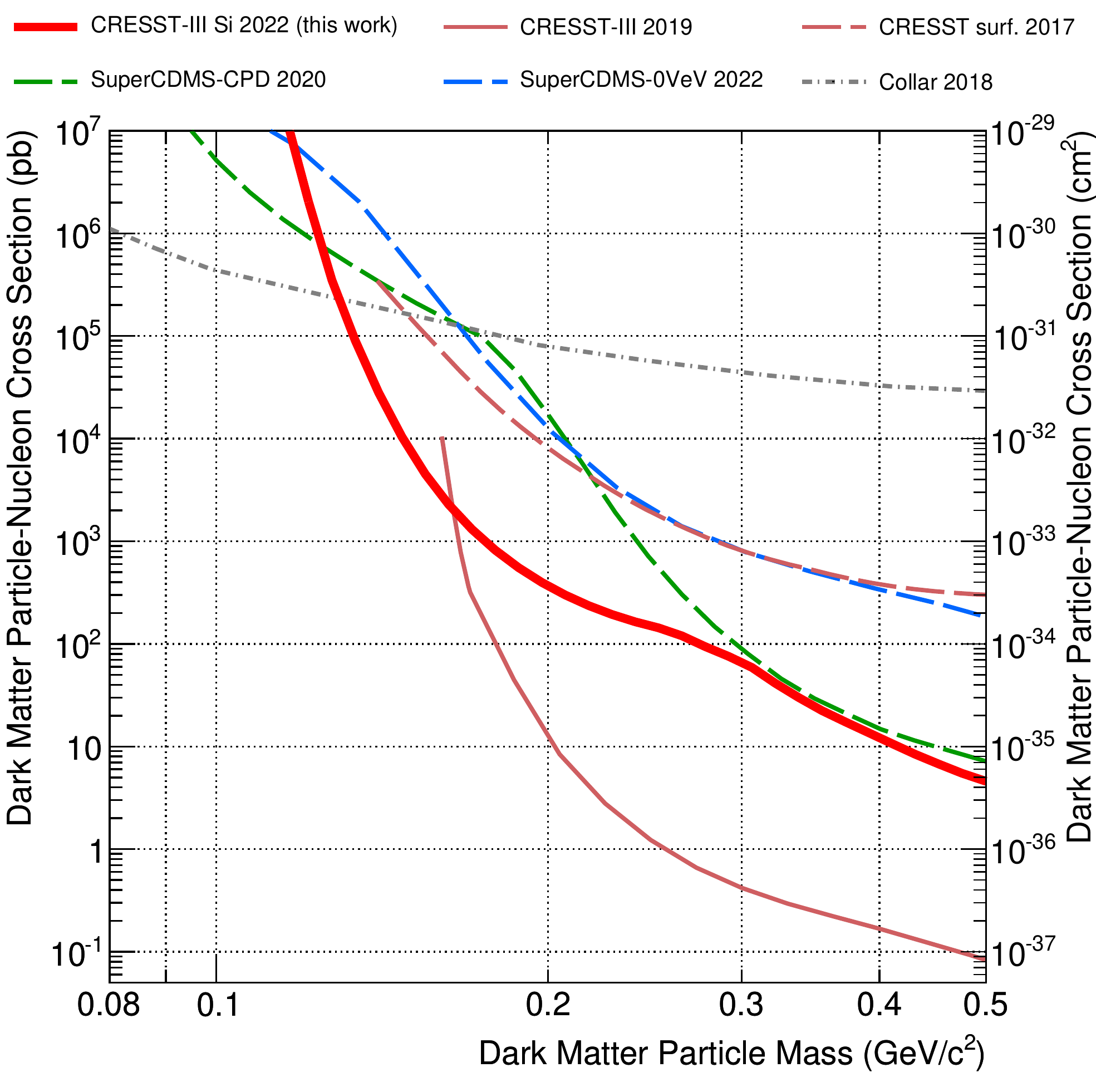}
  \caption{Upper limit on the elastic spin-independent DM-nucleon scattering cross section at 90\% confidence level. The thick solid red line shows the result of this work obtained with a Si detector. The previous CRESST-III result from a CaWO$_4$ crystal operated underground~\cite{detA} is depicted with the darker red solid line. Constraints obtained by the solid-state cryogenic detectors operated above ground are shown by dashed lines: dark red for CRESST-surf with an Al$_2$O$_3$ target ~\cite{cresst-surf}; green for SuperCDMS-CPD~\cite{cpd} and blue for SuperCDMS-0VeV~\cite{SuperCDMS_HVeV_excess2022}, both operating a Si target. Results of hydrogenated organic scintillators by J.~I.~Collar~\cite{collar} are shown as the dash-dotted gray line.}
  \label{fig:limits}
\end{figure}

Fig.~\ref{fig:limits} shows the 90\% confidence level upper limit on the spin-independent elastic DM-nucleon scattering cross section for DM particles with masses from \unit[0.115]{GeV/c$^2$} to \unit[0.5]{GeV/c$^2$} obtained with Yellin's optimum interval algorithm~\cite{yellin,yellin2}. With the use of the Si wafer detector, CRESST-III improves the existing limits for DM masses from \unit[130]{MeV/c$^2$} to \unit[165]{MeV/c$^2$} by a factor of up to 20 compared to previous results. The sensitivity to DM with masses above \unit[165]{MeV/c$^2$} is suppressed compared to the previous CRESST-III DM search with a CaWO$_4$ crystal from 2019 due to the higher LEE rate per kg-day and the lower exposure. For high interaction cross sections, the detector is shielded from DM by the atmosphere and the rock overburden. A conservative estimation of the maximal cross section that can be probed in a deep underground laboratory, such as LNGS, is calculated in Ref.~\cite{SIMP_2018}. Furthermore, we estimate the influence of the shielding in LNGS using the \textsc{verne} package~\cite{verne_code_2018,Kavanagh_SIMP_2018}. Both estimations give an upper boundary for the spin-independent cross section of ${\mathcal{O}(\unit[10^{6}]{pb})}$. Thus, the new parameter space explored in this work is expected to be accessible in the LNGS deep underground laboratory, and not blocked by Earth shielding.

Using the standard DM halo model described above we calculate the expected differential energy spectra for two DM masses, \unit[0.3]{GeV/c$^2$} and \unit[0.5]{GeV/c$^2$}, which are exemplarily shown as the pink and blue line, respectively, in Fig. \ref{fig:LEEfit}. For this, we assumed the cross sections for both DM masses at the limits set in this work. 

Additionally, using Si as a target material opens an opportunity to exploit the $^{29}$Si isotope to probe the spin-dependent interaction of DM with nucleons. We calculated the upper limit on spin-dependent DM interaction with protons and neutrons following the procedure described in~Ref.~\cite{Li_paper} and assuming the nuclear spin structure determined in Ref.~\cite{Ressel1993_nuclearSpin}. The result did not yield an improvement of existing limits on this interaction channel due to the relatively low abundance of 4.6\% of the $^{29}$Si isotope, we will therefore not discuss them in more detail in this work.

\section{Conclusion}
In this work, we have presented the results of a Si detector module of the CRESST-III experiment consisting of a thin wafer, the target, and a larger bulk detector. It features an excellent energy resolution of ${\sigma_{\text{BL}}=\unit[(1.36\pm 0.05)]{eV}}$ and corresponding energy threshold of ${E_{\text{thr}}=\unit[(10.0\pm 0.2)]{eV}}$, set at ${\unit[7.35]{\sigma_{BL}}}$ level. We have probed spin-independent DM-nucleus elastic scattering with DM masses from 115 to \unit[500]{MeV/c$^2$}. For DM masses below \unit[160]{MeV/c$^2$} we improved the existing limits by up to a factor of 20.

The sensitivity of the detector was limited by a sharply rising excess of events at low energies, the LEE. In this work, we have excluded surface backgrounds as a dominant source of this excess by showing that the low energy event rate is not reduced by removing coincidences between the target wafer and the bulk detector. Furthermore, we have shown that noise trigger contributions to the excess are negligible. All the knowledge gained so far in CRESST points to the importance of solid-state physics aspects for the LEE. Therefore, the main focus of the currently on-going studies is to investigate the influence of the material properties of the detector components, the crystal interfaces with holding structures and sensors, and the orientation of the crystal axes on the LEE. In addition, new detector module designs with reduced external stress on the target crystal are being developed and tested.

\section*{Acknowledgments}
We are grateful to LNGS for their generous support of CRESST. 
This work has been funded by the Deutsche Forschungsgemeinschaft (DFG, German Research Foundation) under Germany's Excellence Strategy – EXC 2094 – 390783311 and through the Sonderforschungsbereich (Collaborative Research Center) SFB1258 ‘Neutrinos and Dark Matter in Astro- and Particle Physics’, by the BMBF 05A20WO1 and 05A20VTA and by the Austrian science fund (FWF): I5420-N, W1252-N27, and FG1, and by the Austrian research promotion agency (FFG), project ML4CPD. JB and HK were funded through the FWF project P 34778-N ELOISE. The Bratislava group acknowledges a partial support provided by the Slovak Research and Development Agency (projects APVV-15-0576 and APVV-21-0377). The computational results presented were partially obtained using the Max Planck Computing and Data Facility (MPCDF) and the CBE cluster of the Vienna BioCenter.


\bibliographystyle{apsrev4-2}

\bibliography{apssamp}

\end{document}